\documentstyle[12pt,epsfig]{article}
\input epsf

\newcommand{\be}[1]{
\begin{eqnarray}\label{#1}}
\newcommand{\ee}{\end{eqnarray}}
\newcommand{\ci}[1]{\cite{#1}}
\newcommand{\re}[1]{(\ref{#1})}

\newcommand{\ba}{\begin{array}}
\newcommand{\ea}{\end{array}}
\newcommand{\mpp}{m_{\pi\pi}}
\newcommand{\lo}{{\rm ln}}
 \def\Dirac#1{#1\hskip-6pt/}
\newcommand{\partialboth}{\stackrel{\leftrightarrow}{\partial}}
\newcommand{\chil}[1]{\stackrel{o}{#1}}

\newcommand{\insertfig}[2]{\mbox{\epsfxsize=#1cm \epsfbox{#2.eps}}}

\begin{document}
\rightline{RUB-TPII-02/02}

\begin{center}
{\Large One Loop Chiral Corrections to Hard Exclusive Processes:\\ I. Pion Case}\\[0.5cm]

 N. Kivel$^{a,b}$, M.V.~Polyakov$^{a,c}$\\[0.3cm]

\footnotesize\it $^a$ Petersburg
Nuclear Physics Institute, Gatchina, St. Petersburg 188350,
Russia\\
\footnotesize\it $^b$ Institute for Theoretical Physics ,
Regensburg University, Regensburg, Germany\\
 \footnotesize\it $^c$ Institute for
Theoretical Physics II, Ruhr University Bochum, Germany
\\

\end{center}
\begin{abstract}
We computed the leading non-analytic chiral corrections
to the generalized parton distributions (GPDs) of the pion
and to the two-pion distribution amplitudes. This allows us to
obtain the corresponding corrections for the hard exclusive
processes, such as $\gamma^* \gamma\to \pi\pi$, $\gamma^* N
\to 2\pi N'$
and deeply virtual
Compton scattering on the pion target.
\end{abstract}

\section{Introduction}

Hard processes are known to provide us with valuable information
about the quark and gluon structure of hadrons in terms of parton
distributions and parton distribution amplitudes. The generalized
parton distributions (GPDs) \cite{Mul94,Ji97b,Rad97,Col97}, entering the QCD description of the
hard exclusive processes, interpolate, in a sense,
between usual parton distribution, distribution amplitudes and elastic hadron form factors
(for a review see e.g. \cite{revhard}). GPDs are determined by the low energy physics,
therefore their
dependence on the quark mass, small momentum transfer, etc. can be studied with help
of chiral perturbation theory (ChPT).

In the present paper we develop the ChPT for the simple case of the GPDs in the pion and
two pion distribution amplitudes ($2\pi$DA). We present the
results at the one loop level of the ChPT. In this way we compute the leading non-analytic
corrections of the type $p^2\ \lo(p^2)$ (where $p^2\sim m_\pi^2\sim t$) to GPDs. Such corrections
are universal and allow us to get an insight into structure of the GPDs. Additionally
the leading non-analytic chiral correction to GPDs can be immediately translated to
corresponding correction for the exclusive hard processes such as $\gamma^* \pi \to \gamma \pi$,
$\gamma^* N
\to 2\pi N'$,
$\gamma^*\gamma \to \pi \pi$, etc. Such chiral corrections to the hard exclusive processes are
computed in the present paper for the first time.

\section{Chiral expansion for the light-cone matrix elements}
In this section we discuss the matching of the light-cone quark-gluon operators to
the operators in the effective field theory.

The generalized parton distributions (GPDs) and distributions amplitudes
 are defined as various matrix elements
of the quark-gluon operators on the light cone. Let us introduce
left and right twist-2 quark operators on the light cone:

\be{LRoperators}
\nonumber
O^L_{fg}(\lambda)=
\bar \psi_g\left(\frac{\lambda n}{2}\right)\Dirac{n}\
\frac{1+\gamma_5}{2}\
\psi_f\left(-\frac{\lambda n}{2}\right)\, ,\\
O^R_{fg}(\lambda)=
\bar \psi_g\left(\frac{\lambda n}{2}\right)\Dirac{n}\
\frac{1-\gamma_5}{2}\
\psi_f\left(-\frac{\lambda n}{2}\right)\, .
\ee
Here the vector $n^\mu$ is the light-cone vector $n^2=0$, $f,g$
stand for flavour indices. It is always assumed
the colour gauge link along a straight line between the points $\lambda n/2$
and $-\lambda n/2$. In the effective field theory the operators (\ref{LRoperators})
are matched to the operators formulated in terms of effective
degrees of freedom:

\be{match}
O^{L,R}(\lambda)=F \otimes O_{\rm eff}^{L,R}(\lambda)\, ,
\ee
where $O_{\rm eff}^{L,R}(\lambda)$ is an effective hadronic operator with the same
quantum numbers (but not necessarily with the same twist) as the
quark operators (\ref{LRoperators})
and $F$ is the generating function for the c-number coefficients
which are input for effective field theory. In order to make
sense out of decomposition (\ref{match}) we need to have
systematic power counting rules for construction of the hadronic
operator $O_{\rm eff}^{L,R}(\lambda)$.

As usually we are going to use the Goldstone bosons of
spontaneous chiral symmetry breaking as degrees of freedom for the
construction of the effective operators. The standard power
counting of the chiral perturbation theory (ChPT) uses the fact
that the Goldstone bosons do not interact at zero momentum.
Therefore on the level of this effective field theory, the expansion
amounts to a derivative expansion  of the effective Lagrangian
\cite{Leutw2}
(see \cite{chptrev} for introduction to ChPT). The lowest
order term  reads:
\be{ChL} {\cal L}_{eff}={\cal L}_2\,+\,
...=\frac{F_\pi^2}{4}\langle
\partial_\mu U\partial^\mu U^\dagger+\chi U^\dagger+\chi^\dagger U
\rangle\, +\, ...
\ee
where the Goldstone boson fields $U(x)=\exp(i\pi^a(x) \tau^a/F_\pi)$,
$\chi=2B{\rm diag}(m_u, m_d)$\footnote{We use standard notation for quark
condensate $\langle\bar\psi\psi\rangle=-F^2_\pi B+O(p^2 \ \lo p)$} and dots
denote higher order terms. We consider two flavour case and
do not introduce external fields for brevity.
The fields $U, \chi$ that occur in
the effective Lagrangian are subject to the following chiral counting rules:
\be{Chcount}
U\, \sim\, O(p^0),\quad
\partial_\mu U\, \sim\, O(p^1),\quad \chi  \sim\, O(p^2)
\ee
where  $p$ is small momentum, i.e. small parameter of
chiral expansion. From \re{Chcount} one finds that
 the leading order effective Lagrangian \re{ChL} is of order $p^2$.
We would like to emphasize, that locality  together with
condition $UU^\dagger=1$ plays  important role  in derivation
of the chiral expansion \ci{Leutw2}.

On the other hand, the description of the many hadron hard
reactions grounds on the QCD collinear factorization. In  such
approach the non-perturbative part associated with a soft physics
is parametrized by the matrix elements of some non-local
light-cone operators. These objects appear as natural constructing
blocks and it is convenient to keep these  operators non-local
without transition to a series of the local operators. Such
approach is useful in the higher energy phenomenology and we would
like to follow this philosophy in the effective theory.

Therefore, to perform the matching \re{match} in terms of the
effective fields  the standard counting rules \re{Chcount} should
be slightly extended. The point is that although after the QCD
factorization the soft part of the hard processes does not contain
the hard momenta, it still ``remembers'' about them. In
the operators \re{LRoperators} such ``memory'' is reflected by the
dependence on the light-cone vector $n^\mu$ and we have to specify
the chiral order of this parameter.

Let us mention that the light-cone decomposition of any four-vector
$V^\mu$ reads:
\be{sudakov}
V^\mu=V^+\ \widetilde n^\mu + V^-\ n^\mu +V_\perp^\mu\, .
\ee
Here are $n^\mu$ and $\widetilde n^\mu$ are light-cone vectors
$n^2=\widetilde n^2=0$
which we normalize as $n\cdot \widetilde n=1$. These two vectors
define two-dimensional plane, the perpendicular plane is called
transverse plane. The vectors from the transverse plane
$V_\perp^\mu$ by definition satisfy $n\cdot V_\perp=\widetilde n\cdot
V_\perp=0$. The physical observables are obviously invariant under
rescaling of the vector $n$, $i.e.$ under transformation $n^\mu\to c\ n^\mu$
where $c$ is an arbitrary nonzero constant.
This invariance corresponds to the boost invariance of the physical observables.
It is convenient  to fix
the normalization of the light-cone vector $n^\mu$ by condition
like $n\cdot p=1$ where $p$ is one of the small external momenta
entering the soft part of the amplitude.
Such condition implies that the light-cone vectors $n^\mu\sim O(1/p)$
and $\widetilde n^\mu \sim O(p)$, where $p$ assumed to be a generic
soft momentum as in power counting \re{Chcount}.

To summarize, we have to construct the effective hadronic
operator in eq.~(\ref{match}) using as building blocks chiral
fields $U(x)$ and their derivatives with counting rules:
\be{count}
n\cdot \partial U(x)\sim O(p^0), \quad \widetilde n\cdot \partial U(x)\sim
O(p^2),\quad \partial_\perp U(x)\sim O(p)\, .
\ee
Using these building blocks one can derive that
in effective field theory the operators (\ref{LRoperators}) are
matched to the operators in terms of Goldstone
degrees of freedom with the same quantum numbers:
\be{matching}
\nonumber
O^L_{fg}(\lambda)&=&
\frac{i F_\pi^2}{4}\ \int_{-1}^1 d\beta \int_{-1+|\beta|}^{1-|\beta|} d\alpha\
F(\beta , \alpha)
 \left[U\left(\frac{\alpha+\beta}{2}\lambda n\right)n\cdot\partialboth
U^\dagger\left(\frac{\alpha-\beta}{2}\lambda n\right)\right]_{fg} +\ldots\, ,\\
O^R_{fg}(\lambda)&=&
\frac{i F_\pi^2}{4}\ \int_{-1}^1 d\beta
\int_{-1+|\beta|}^{1-|\beta|} d\alpha\ F(\beta , \alpha)
 \left[U^\dagger\left(\frac{\alpha+\beta}{2}\lambda n\right)n\cdot\partialboth
U\left(\frac{\alpha-\beta}{2}\lambda n\right)\right]_{fg} +\ldots
\ee
Here $F(\beta , \alpha)$ is the generating function of the tower of
low-energy constants and $\partialboth$ denotes
$\stackrel{\rightarrow}{\partial}-\stackrel{\leftarrow}{\partial}$.
The low-energy constants are characteristics
of the structure of the pion, they are not determined in the effective field theory.
The ellipsis in eqs.~(\ref{matching})
stands for operators which do not contribute to the one and two pion matrix
elements of the operators $O^{L,R}$ or which are
of higher orders in the chiral counting. Note that if one would
consider the chiral corrections say for three pion distribution
amplitudes one would need to add additional operators to
eq.~(\ref{matching}).
In the next section
we consider chiral expansion of some light-cone
matrix elements at the leading order using  formulae \re{matching}.

Let us note that for the gluon light-cone operators the construction of
the corresponding effective operators
is actually the same as for the singlet quark operators.
The only difference that for gluon operators
we have to introduce an independent generating function $F_G(\beta,\alpha)$.
We do not discuss the gluon distributions
in the present paper because their discussion repeats
almost word by word that for the
singlet quarks distributions.

\section{Light-cone matrix elements in the leading order of chiral
expansion}
\begin{figure}[t]
\unitlength1mm
\begin{center}
\hspace{0cm}
\insertfig{4}{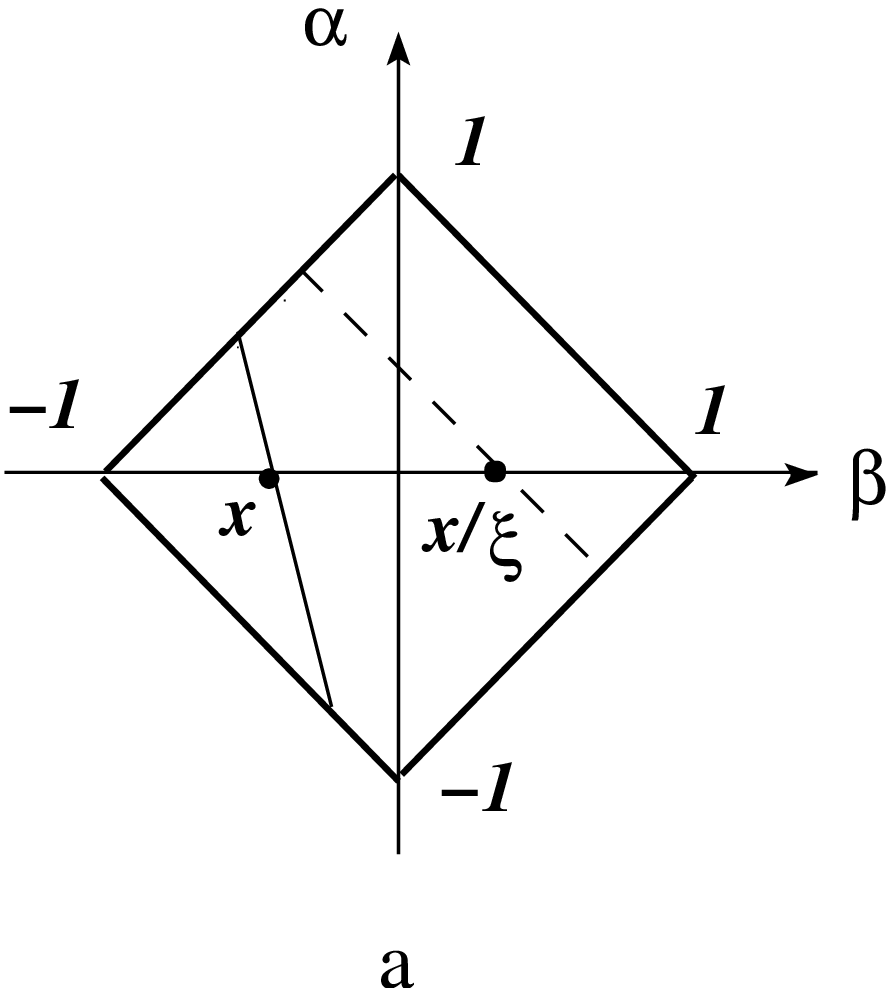}
\hskip1cm
\insertfig{4}{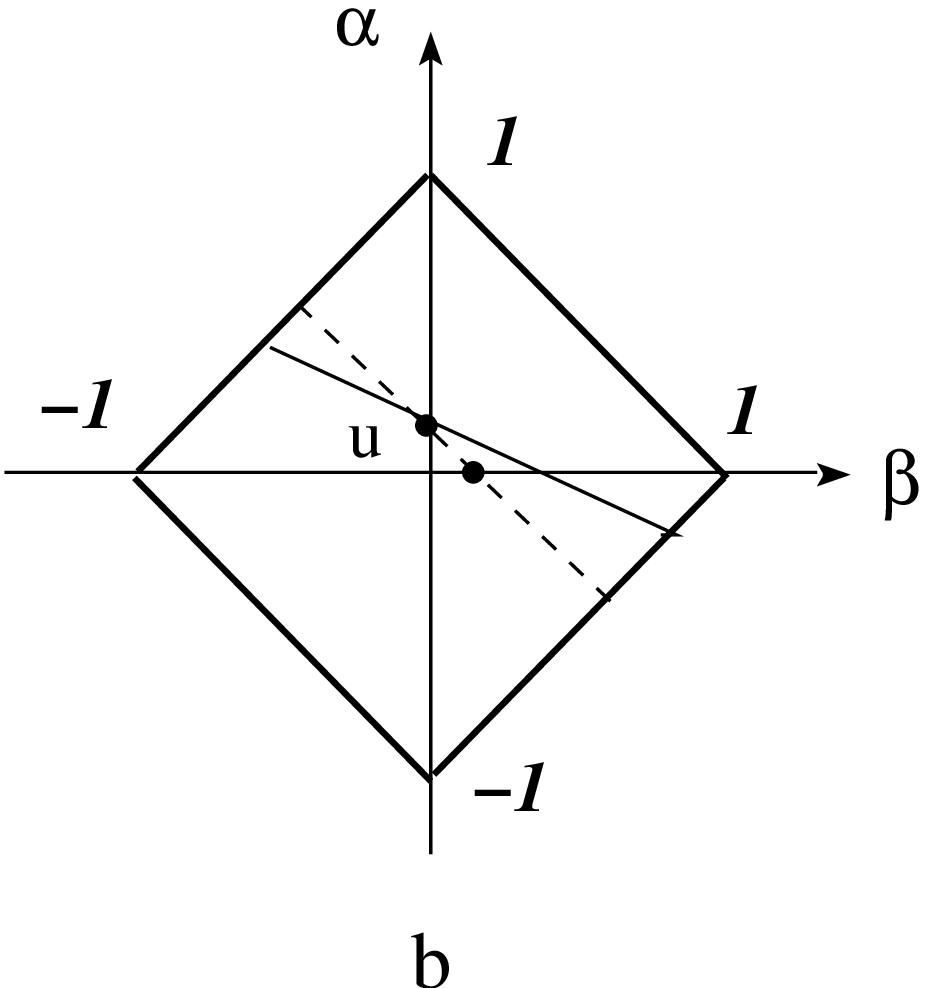}
\end{center}
\vspace{-0.5cm}
\caption[dummy]{\small
Integration lines for integrals relating GPDs ${H^{I}}(x,\xi)$ (a)
and 2$\pi$DAs $\Phi^{I}_{2\pi}(u,\eta)$ (b) with DD $F^I(\beta,\alpha)$.
Dashed line is a typical  integration line for the D-term.
\label{rombus}}
\end{figure}

In order to see the physical meaning of the generating function
$F(\beta , \alpha)$ in \re{matching} we compute the GPDs of the pion
and $2\pi$DAs in the leading order of chiral expansion. GPDs of the
pion are defined as:
\be{GPDdef}
\int \frac{d \lambda}{2\pi}\ e^{-i (n\cdot P) x \lambda}
\langle\pi^b(p')|{\rm tr}\left[\hat T\ O^{L+R}(\lambda)
\right]|\pi^a(p)\rangle=
\left\{
\begin{array}{cr}
2i \varepsilon^{abc}\ H^{I=1}(x,\xi,t),\ {\rm for}\ \hat T=\tau^c \\
  2\delta^{ab}\ H^{I=0}(x,\xi,t),\ {\rm for}\ \hat T=1
\end{array}
\right.
\ee
Here we introduced the nonsinglet $H^{I=1}$ and singlet $H^{I=0}$ GPDs of the pion.
In what follows
we use the standard notations for kinematical variables:
$P=\frac 12 (p+p')$, $\xi=-\frac{n\cdot(p'-p)}{n\cdot (p+p')}$
and $t=(p'-p)^2$. Now substituting the expressions for the light-cone operators
in the effective theory (\ref{matching})
into the definition of GPDs (\ref{GPDdef}) and computing the corresponding matrix
elements at the tree level
we obtain the following expressions for GPDs in the leading order of ChPT
(denoted $\chil{H}$):
\be{tree}
\chil{H^{I}}(x,\xi)= \int [d\alpha d\beta]\ F^{I}(\beta,\alpha)\
\left[\delta(x-\xi \alpha-\beta)-(1-I)\ \xi \delta(x-\xi (\alpha+\beta))
 \right]\, .
\ee
Here we introduce the notations
$F^{1,0}(\alpha,\beta)=\frac{1}{2}(F(\beta , \alpha)\pm F(-\beta,\alpha))$
and $[d\alpha d\beta]$ stands for the integration over the rombus $|\alpha|+|\beta| \leq 1$,
see Fig.~\ref{rombus}.
In the first term of eq.~(\ref{tree}) one recognizes immediately the double distribution representation
for the GPD \cite{Mul94,Rad98}. The second term\footnote{which contributes
only for the singlet GPD} in eq.~(\ref{tree}) corresponds to the D-term contribution
\cite{Pol99b} to the GPDs. Note that the D-term for the pion GPD is also fixed in terms
of the double distribution
due to the soft pion theorem of ref.~\cite{Pol99a}

\be{spt1}
\chil{H}{}^{I=0}(x,\xi=\pm1)=0.
\ee

{}From this simple exercise we learned that the generating
function $F(\beta , \alpha)$ for the low-energy chiral constants
coincides in the leading order of ChPT with
the double distribution for the pion in the chiral limit and at zero momentum transfer squared.
This implies that the function $F(\beta,\alpha)$ is related to
the quark distributions in the pion in the chiral limit ($m_\pi=0$):

\be{Squarks}
\int_{-1+|\beta|}^{1-|\beta|}d\alpha\ F^{I=0}(\beta , \alpha)=
\frac12\left[\theta(\beta)
 \chil{q}(\beta)-\theta(-\beta)\chil{ \bar q}(-\beta)\right],
\\
\int_{-1+|\beta|}^{1-|\beta|}d\alpha\ F^{I=1}(\beta , \alpha)=
\theta(\beta)
 \chil{q}(\beta)+\theta(-\beta)\chil{\bar q}(-\beta)\, .
\ee
The first moment of these distributions is related to the forward matrix elements of the
energy momentum tensor and vector current respectively. This gives:
\be{srules}
 \int [d\alpha d\beta]\ F^{I=0}(\beta,\alpha)\beta &=& M_2^Q\, \\
 \int [d\alpha d\beta]\ F^{I=1}(\beta,\alpha) &=& 1\, .
\ee
where  we introduce notation for the fraction of the pion momentum
carried by quarks and antiquarks
$M_2^Q=\int_0^1 dx\ x\ (q(x)+\bar q(x))$.
Using these equations and \re{tree} one can easily obtain first
moment for GPD $H^{I=1,0}$:
\be{}
\int_{-1}^{1}dx \, x\chil{H}{}^{I=0}(x,\xi)& =& (1-\xi^2)M_2^Q\, \\
\int_{-1}^{1}dx \, \chil{H}{}^{I=1}(x,\xi) &=&1\, .
\ee

The interpretation of the generating functions $F^{I}(\beta,\alpha)$ as DD's
means that these functions depend on the factorization scale $\mu$.
The functional dependence from this parameter is described by the evolution
equations \cite{Rad97}. For the sake of simplicity we do not write this
argument explicitly but imply it.

Because the matching of the operators (\ref{LRoperators}) is universal,
the same effective operators (\ref{matching}) can be used to define
the pion and two-pion distribution amplitudes in the leading order of ChPT.

The pion DA is defined as:
\be{pi}
\int \frac{d\lambda}{2\pi} e^{-i(n\cdot p) u\lambda}
 \langle \pi^a(p) | {\rm tr} \left[\tau^b
\ O^{L-R}(\lambda)\right]|0\rangle =
i\delta^{ab}F_\pi\ \phi_\pi(u)\, .
\ee
In the leading order of ChPT we obtain for the pion DA
in the chiral limit (refereed to as $\chil{\phi}_\pi$):

\be{pionDAtree}
\chil{\phi}_\pi(u)=\int [d\alpha d\beta]\ F^{I=1}(\beta,\alpha)\
\delta(u- \alpha-\beta)\, .
\ee
Comparing this expression with eq.~(\ref{tree}) we recover that
\be{sptH2}
\chil\phi_\pi(u)=\pm\chil{H}{}^{I=1}(u,\xi=\pm1)\, .
\ee
This is again in an agreement with the soft pion theorem
of ref.~\cite{Pol99a}.

The two-pion DAs ($2\pi$DAs) are defined as:
\be{2piDAsdef}
\int \frac{d \lambda}{2\pi}\ e^{-i u \lambda (n\cdot (p+p'))/2}
\langle \pi^a(p)\pi^b(p')|
{\rm tr}\left[\hat T\ O^{L+R}(\lambda) \right]|0\rangle=
\left\{
\begin{array}{cr}
 i \varepsilon^{abc}\ \Phi_{2\pi}^{I=1}(u,\eta,\mpp),\ {\rm for}\
\hat T=\tau^c \\
\delta^{ab}\ \Phi_{2\pi}^{I=0}(u,\eta,\mpp),\
{\rm for}\ \hat T=1
\end{array}
\right.
\ee
with the kinematical variables $\eta=\frac{n\cdot (p'-p)}{n\cdot(p+p')}$ and
$\mpp^2=(p+p')^2>0$.
The leading ChPT order expression for $2\pi$DAs (referred to as $\chil{\Phi}_{2\pi}$) has the form:
\be{DAtree}
\chil{\Phi}{}^{I}_{2\pi}(u,\eta)= \int [d\alpha d\beta]\ F^{I}(\beta,\alpha)\
 \left[\eta\ \delta(u-\alpha-\eta \beta)-(1-I)\  \delta(u-\alpha-\beta))
 \right]\, .
\ee
We see that the $2\pi$DAs are expressible in terms of the same double distribution.
In Fig.~\ref{rombus}
we illustrated how to obtain pion GPDs, D-term and pion and $2\pi$DAs from the double distribution $F(\beta , \alpha)$.
{}From eq.~(\ref{srules}) we obtain
\be{2piSR}
 \int_{-1}^{1}du \,u \chil{\Phi}{}^{I=0}_{2\pi}(u,\eta)  &=& -(1-\eta^2)M_2^Q\, \\
\int_{-1}^{1}du \, \chil{\Phi}{}^{I=1}_{2\pi}(u,\eta) &=& 1\, .
\ee
Again, using \re{DAtree} and \re{pionDAtree} one easily obtain soft pion theorems
\cite{Pol99a,Teryaev}:
\be{sptDA}
\chil{\Phi}{}^{I=0}_{2\pi}(u,\eta=\pm 1)&=&0\, ,\\
\chil{\Phi}{}^{I=1}_{2\pi}(u,\eta=\pm 1)&=&\pm \chil{\phi}_\pi(u)\, .
\ee

\section{Leading non-analytic chiral corrections to GPDs and DAs}
In what follows we consider
the next-to-leading chiral corrections of the type $p^2 \lo(p^2)$ with
$m_\pi^2\sim t\sim \mpp^2\sim p^2$. These leading non-analytic
contributions to the pion
GPDs and DAs are universal in the sense that they are
expressible completely in terms of the leading order generating function
$F(\beta , \alpha)$. The corresponding corrections are obtained computing
the loop diagrams shown on Figs.~\ref{diags},\ref{DAdiags}. The
divergencies
contained in the one loop graphs can be removed by renormalization of the
coupling constants which appear in the NLO of ChPT. Then loop contributions
depends on the chiral renormalization scale $\mu_{\chi}$. A change in this scale however
only adds a constants which will shift running couplings. Because we compute
only the leading non-analytic terms the detailed discussion of this subject is outside
of our consideration.

\subsection{Generalized Parton Distributions}

\begin{figure}[t]
\unitlength1mm
\begin{center}
\hspace{0cm}
\insertfig{12}{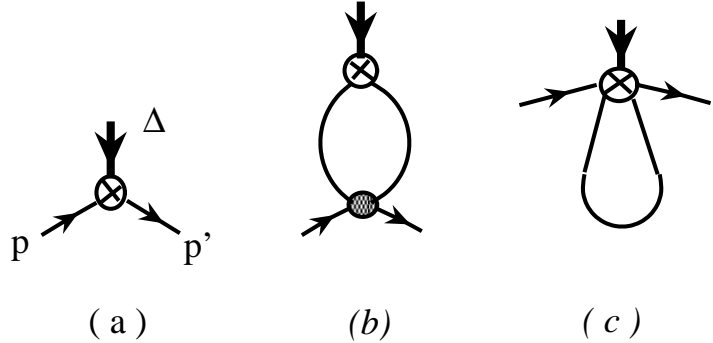}
\end{center}
\vspace{-0.5cm}
\caption[dummy]{\small  Diagrams contributing to GPD matrix elements.
Arrows denote the directions of external momenta.
Tree level (a) and loop corrections {\it (b), (c)}. The four pion vertex in
the diagram {\it (b)} originates  from expansion of effective Lagrangian
\re{ChL}.
\label{diags}}
\end{figure}

Computing  diagrams on Fig.\ref{diags} for the case of pion GPDs we obtain the following
expressions for them with leading non-analytic corrections included:

\be{loopI0}
H^{I=0}(x,\xi,t)&=&\chil{H}{}^{I=0}(x,\xi)\\
&+&
\frac{\theta[|x|\leq \xi]} {2 (4\pi F_\pi)^2}\
[ m_\pi^2-2t]\
\int_{-1}^1 d\eta\  \lo\left[m_\pi^2-(1-\eta^2)\frac{t}{4}\right]
\chil{\Phi}{}^{I=0}_{2\pi}\left(\frac x\xi,\eta\right)
\nonumber
\ee
\be{loopI1}
H^{I=1}(x,\xi,t)&=&\chil{H}{}^{I=1}(x,\xi)\ \left(1-\frac{m_\pi^2 \lo m_\pi^2}{(4\pi F_\pi)^2} \right)\\[-0.2cm]
&+& \frac{ \theta[|x|\leq \xi]}{2 (4\pi F_\pi)^2\xi}
\int_{-1}^1 d\eta
\left[m_\pi^2-(1-\eta^2)\frac{t}{4}\right]\ \lo\left[m_\pi^2-(1-\eta^2)\frac{t}{4}\right]
 \frac{\partial}{\partial\eta}\chil{\Phi}{}^{I=1}_{2\pi}\left(\frac x\xi,\eta\right)
\, .
\nonumber
\ee
In both equations $\chil{H^I}$ and $\chil{\Phi}{}^{I}_{2\pi}$ are the leading
order expressions given by  eq.~(\ref{tree}) and eq.~(\ref{DAtree}) correspondingly. For convenience we do not write explicitly the dependence on $\mu_{\chi}$
assuming it in all logarithms:
\be{much}
\lo\left[m_\pi^2-(1-\eta^2)\frac{t}{4}\right]\equiv
\lo\left[m_\pi^2/\mu_{\chi}^2\right]+\lo\left[1-(1-\eta^2)\frac{t}{4m_\pi^2}\right]
\ee

Nontrivial contributions with two pion DAs $\chil{\Phi}{}^{I}_{2\pi}$
originated  from the
diagram  Fig.~\ref{diags}(b). Appearance of two pion DAs is a direct
consequence of the presence of the two pion state in the  $t$-channel. From  eq.~\re{loopI0} and
eq.~(\ref{loopI1}) we see that in both cases the leading non-analytic in $t$ chiral corrections are nonzero
only in the so-called ERBL-region $|x|\leq \xi$. This shows that
the $t$-dependence of GPDs can not be reduced to popular factorized ansatz $H(x,\xi,t)=H(x,\xi)F(t)$.

At the points
$x=\pm \xi$ chiral corrections nullifies (except trivial $t$-independent term  in the isovector GPD),
because it is expected that the two-pion distribution amplitudes nullify at the end points:
\be{endp}
\chil{\Phi}{}^{I}_{2\pi}(u=\pm 1,\eta)=0.
\ee
Such  behaviour at the points $x=\pm \xi$ is very  important
for validity  of the factorization theorem in different hard exclusive
reactions.

Let us mention that DD's $F^{I}(\beta,\alpha)$ depend on the factorization
 scale $\mu$.
From physical point of view, the evolution in $\mu$ and computing
chiral corrections are independent operations
and hence must commute with each other.
Using formulae \re{loopI0} and \re{loopI1} one can easily see that this
property takes place. Consider, for simplicity,
the leading logarithmic approximation for the evolution in factorisation scale
$\mu$.
To this accuracy one can construct multiplicatively renormalizable moments:
\be{Gmom}
H_n^{I}(\xi,t)& =&\int_{-1}^{1}dx\ H^{I}(x,\xi,t)\ C^{3/2}_n(x/\xi)\\
\Phi_n^{I}(\eta,\mpp)& =&\int_{-1}^{1}du\ \Phi^{I}_{2\pi}
(u,\eta,\mpp)\ C^{3/2}_n(u)
\ee
where $C^{3/2}_n(u)$ is a Gegenbauer polynomial. Their evolution is given
by simple equations\footnote{We consider only isovector case I=1 for simplicity}:
\be{evolGPD}
H_n^{I=1}(\xi,t|{\mu^2})&=& L(\mu,\mu_0)^{\gamma_n/\beta_0}H_n^{I=1}
(\xi,t|{\mu^2_0})
\\  \label{evolDA}
\Phi_n^{I=1}(\eta,\mpp |{\mu^2})& =&
L(\mu,\mu_0)^{\gamma_n/\beta_0}\Phi_n^{I=1}(\eta,\mpp |{\mu^2_0})
\ee
where $L(\mu,\mu_0)=\alpha(\mu^2)/\alpha(\mu^2_0)$, $\gamma_n$ and
$\beta_0$ is leading order anomalous dimension and QCD $\beta$-function respectively.

Computing the Gegenbauer moments of  eq.~\re{loopI1} we obtain:
\be{ChPTmom}
H_{2n}^{I=1}(\xi,t|{\mu^2})&=& \chil{H}{}^{I=1}_{2n}(\xi|{\mu^2})
\left(1-\frac{m_\pi^2 \lo m_\pi^2}{(4\pi F_\pi)^2} \right)+
\\&&
\frac{1} {2 (4\pi F_\pi)^2}
\int_{-1}^1 d\eta\
\left[m_\pi^2-(1-\eta^2)\frac{t}{4}\right]\ \lo\left[m_\pi^2-(1-\eta^2)\frac{t}{4}\right]
\frac{\partial}{\partial\eta}
\chil{\Phi}{}^{I=1}_{2n}\left(\eta|\mu^2 \right)\, . \nonumber
\ee
Using \re{evolGPD} and \re{evolDA} we obtain that above equation remains unchanged.
In other words, chiral corrections are decoupled from evolution as it should be.

Chiral loops generate corrections to the soft pion theorem
 \re{spt1}. Taking limit $\xi\to \pm 1,\, t\to 0$ we obtain:
\be{NLO0}
 H^{I=0}(x,\xi=\pm 1,t=0)&=&
\frac{ m_\pi^2\lo[m_\pi^2] }{2 (4\pi F_\pi)^2}
\int_{-1}^1 d\eta\
\chil{\Phi}{}^{I=0}_{2\pi}\left(x,\eta\right)\, .
\ee
Whereas the soft pion theorem (\ref{sptH2}) does not get
corrections of the order $m_\pi^2 \lo(m_\pi)$
\be{NLO1}
H^{I=1}(x,\xi=\pm 1,t=0)&=&
\chil{\phi}_\pi(u)\ +O(m_\pi^2)\, .
\ee

Now we consider various limiting cases of the
obtained results
for the chiral expansion of the pion GPDs.

\vspace{0.2cm}
\noindent
\underline{\em Sum rules}\\[0.2cm]
The first Mellin moment of the non-singlet GPD $H^{I=1}$ is the related to the pion
electromagnetic form factor

\be{norm}
\int_{-1}^1 dx\ H^{I=1}(x,\xi,t)=F^{\rm e.m.}_\pi(t)\,.
\ee
Integrating the the one-loop result (\ref{loopI1}) we obtain the well
known result for the leading non-analytic contribution to the
pion e.m. form factor\cite{Leutw2}

\be{ffr}
F^{\rm e.m.}_\pi(t)=1-\frac{m_\pi^2 \lo m_\pi^2}{(4\pi F_\pi)^2}+
\frac{1}{ (4\pi F_\pi)^2}
\int_{0}^1 d\eta
\left[m_\pi^2-(1-\eta^2)\frac{t}{4}\right]\
\lo\left[m_\pi^2-(1-\eta^2)\frac{t}{4}\right]\,.
\ee
The second moment of the singlet GPD $H^{I=0}$ is related to the
form factors of  quark part of the energy momentum tensor:

\be{1mom}
\int_{-1}^1 dx\ x\ H^{I=0}(x,\xi,t)=2\ \theta_2(t)- 2\ \xi^2\
\theta_1(t)\,,
\ee
where $\theta_{1,2}(t)$ are the pion form factors of
the quark part of the energy momentum
tensor \cite{Pag66,Don91}:

\be{emt}
\langle p'| T^{\mu\nu}_Q|p\rangle=2\ P^\mu P^\nu \theta_2(t)+\frac
12 (g^{\mu\nu}\Delta^2-
\Delta^\mu \Delta^\nu) \theta_1(t) \, .
\ee
Computing the second Mellin moment of our results for the singlet
GPD (\ref{loopI0}) we obtain the leading non-analytic corrections
for the energy momentum tensor:

\be{EMTFF}
\theta_2(t)&=& \frac 12\ M_2^Q \left(1+ O(p^2) \right)\, \\
\theta_1(t)&=& \frac 12\ M_2^Q \left(1+
\frac{m_\pi^2-2 t}{(4\pi F_\pi)^2}
\int_{0}^1 d\eta \
(1-\eta^2)\
\lo\left[m_\pi^2-(1-\eta^2)\frac{t}{4}\right]
 \right)\, ,
\ee
This result coincides with that of ref.~\cite{Don91}.

\vspace{0.2cm}
\noindent
\underline{\em Forward limit and parton distributions in the transverse plane}\\[0.2cm]
In the forward limit, $i.e.$ $\xi\to 0, t\to 0$ the GPDs $H^{I=0}$
and $H^{I=1}$ are reduced to the singlet and non-singlet quark
distributions in the pion correspondingly. It is easy to see from eq.~(\ref{loopI0}) that the singlet
quark distribution in the pion does not receive non-analytical
chiral corrections. The same conclusion was reached in
refs.~\cite{Arn01,Che01}. As to nonsinglet quark distribution, the
forward limit of eq.~(\ref{loopI1}) gives:

\be{zero}
q(x)=\ \chil{q}(x) \left(1-\frac{m_\pi^2 \lo m_\pi^2}{(4\pi
F_\pi)^2}\right)+ \delta(x)\ \frac{m_\pi^2 \lo m_\pi^2}{(4\pi
F_\pi)^2}\, .
\ee
This is exactly the result obtained recently in
refs.~\cite{Arn01,Che01}.
Note that two limits $\xi\to 0$ and  $m_\pi\to 0$ do not commute
The above result is obtained by
taking the limit $\xi\to 0$ {\em before} the limit $m_\pi\to 0$.

Let us now consider more general limit  when $\xi\to 0$ but $t\neq
0$. In such limit the GPDs can interpreted as
the Fourier transform of
the probability
distribution of partons in the transverse plane \cite{Burk01}.
Taking this particular limit we again obtain that the singlet GPDs has
no the non-analytic chiral corrections, whereas the result for the
the non-singlet GPDs can be written as:

\be{qt}
\nonumber
H^{I=1}(x,0,t)&\equiv& q(x,t)=\ \chil{q}(x) \left(1-\frac{m_\pi^2 \lo m_\pi^2}{(4\pi
F_\pi)^2}\right)\\
 &+& \delta(x)\
\frac{1}{ (4\pi F_\pi)^2}
\int_{0}^1 d\eta
\left[m_\pi^2-(1-\eta^2)\frac{t}{4}\right]\
\lo\left[m_\pi^2-(1-\eta^2)\frac{t}{4}\right]\,.
\ee
The probability
distribution of partons in the transverse plane is obtained from the
above result by the Fourier transformation \cite{Burk01}:

\be{prob}
f(x,b_\perp)=\int \frac{d^2 \Delta_\perp}{(2\pi)^2}\
H(x,0,-\Delta_\perp^2)\ e^{i\ b_\perp\cdot \Delta_\perp}\, .
\ee
As the expression (\ref{qt}) is valid only for small values of $t$
we are able to derive only the behaviour of $f(x,b_\perp)$ at
large values of the impact parameter $b_\perp$:

\be{eq1}
f(x,b_\perp) &\sim& \delta(x)\
\frac{1}{(4\pi F_\pi)^2}\ \frac{2}{3 \pi b_\perp^4}\, \quad {\rm
for}\quad \frac{1}{4 \pi F_\pi} \ll b_\perp \ll \frac{1}{m_\pi}\,.
\ee
{}From eq.~(\ref{qt}) we can also obtain the leading chiral
contribution to the average width of the $b_\perp$ distribution
\be{eq2}
\langle b_\perp^2 \rangle=
\int d^2b_\perp\ b_\perp^2\ f(x,b_\perp)=-\frac 23\ \delta(x)\
\frac{\lo\ m_\pi^2}{(4\pi F_\pi)^2}\, .
\ee
We see that the width of the parton distribution in the transverse
plane is divergent in the chiral limit. This divergent piece
arises from the contribution of the long-range pion cloud. Note
that partons which are ``responsible" for the divergent chiral
contribution are ``concentrated" in the region of small
\footnote{More precisely at $x\ll m_\pi^2/(4\pi F_\pi)^2$}
$x$.

\vspace{0.2cm}
\noindent
\underline{\em Extrapolation to $t=0$}\\[0.2cm]
Generalized parton distributions can be probed in the hard
exclusive reactions. Usually the point $t=0$ is not accessible
directly in these processes. Our results for the leading
non-analytical corrections for the GPDs allow us to study the
extrapolation of the GPDs to the point $t=0$. Exactly at $t=0$ the expression
for the pion GPDs with leading non-analytic chiral corrections included has the form:

\be{loopI0zero}
H^{I=0}(x,\xi,t=0)&=&\chil{H}{}^{I=0}(x,\xi)\\
&+&
\frac{\theta[|x|\leq \xi]} {2 (4\pi F_\pi)^2}\
 m_\pi^2\ \lo(m_\pi^2)
\int_{-1}^1 d\eta\
\chil{\Phi}{}^{I=0}_{2\pi}\left(\frac x\xi,\eta\right)
\nonumber
\ee
\be{loopI1zero}
H^{I=1}(x,\xi,t=0)&=&\chil{H}{}^{I=1}(x,\xi)\ \left(1-\frac{m_\pi^2 \lo m_\pi^2}{(4\pi F_\pi)^2} \right)\\
&+& \frac{ \theta[|x|\leq \xi]}{ (4\pi F_\pi)^2\xi}\  m_\pi^2\ \lo(m_\pi^2)
\chil{\phi}{}_{\pi}\left(\frac x\xi\right)
\, .
\nonumber
\ee
Also from eqs.~(\ref{loopI0},\ref{loopI1}) we can easily obtain
the leading non-analytic contributions to the ``slopes" of the $t$-dependence of
the GPDs. The result is:

\be{loopI0derivzero}
\left.
\frac{\partial}{\partial t}H^{I=0}(x,\xi,t)\right|_{t=0}&=&-
\frac{\theta[|x|\leq \xi]} {(4\pi F_\pi)^2}\
 \lo(m_\pi^2)
\int_{-1}^1 d\eta\
\chil{\Phi}{}^{I=0}_{2\pi}\left(\frac x\xi,\eta\right) \, ,\\
\left.
\frac{\partial}{\partial t}H^{I=1}(x,\xi,t)\right|_{t=0}&=&-
\frac{\theta[|x|\leq \xi]} {4 (4\pi F_\pi)^2\xi}\
 \lo(m_\pi^2)
\int_{-1}^1 d\eta\ \eta\ \chil{\Phi}{}^{I=1}_{2\pi}\left(\frac
x\xi,\eta\right)\, . \nonumber \ee Such relations can be useful
for extrapolation of the experimental data from nonzero $t$ to
the point $t=0$. Also these relations can be useful for
interpretation of the lattice results.

\subsection{Two pion Distribution Amplitudes }
\begin{figure}[t]
\unitlength1mm
\begin{center}
\hspace{0cm}
\insertfig{12}{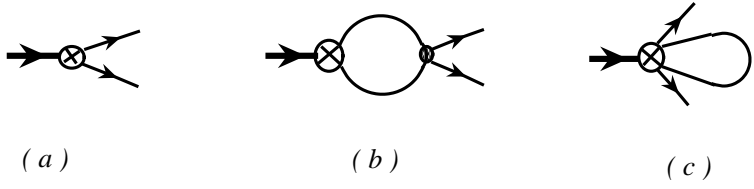}
\end{center}
\vspace{-0.5cm}
\caption[dummy]{\small  Diagrams contributing to two pion DAs
 matrix elements. Tree level {\it(a)} and loop corrections {\it (b), (c)}
\label{DAdiags}}
\end{figure}

Diagrams  for the chiral expansion of the $2\pi$DAs are depicted in
Fig.\ref{DAdiags}. Their contribution reads:
\be{loopDA0}
\nonumber
\Phi_{2\pi}^{I=0}(u,\eta,\mpp)&=&\chil{\Phi}{}^{I=0}_{2\pi}(u,\eta)
+  \frac{1}{2 (4\pi F_\pi)^2}\\
&\times& [m_\pi^2-2\mpp^2]\
\int_{-1}^1 d\eta'\  \lo\left[m_\pi^2-(1-\eta'^2)\frac{\mpp^2}{4}\right]
\chil{\Phi}{}^{I=0}_{2\pi}\left(u,\eta'\right)\, ,
\ee
\be{loopDA1}
&&\Phi_{2\pi}^{I=1}(u,\eta,\mpp)= \chil{\Phi}{}^{I=1}_{2\pi}(u,\eta)\
\left(1-\frac{m_\pi^2 \lo m_\pi^2}{(4\pi F_\pi)^2} \right)\\
&& + \frac{ \eta}{2 (4\pi F_\pi)^2}
\int_{-1}^1 d\eta'
\left[m_\pi^2-(1-\eta'^2)\frac{\mpp^2}{4}\right]\ \lo\left[m_\pi^2-(1-\eta'^2)\frac{\mpp^2}{4}\right]
 \frac{\partial}{\partial\eta'}\chil{\Phi}{}^{I=1}_{2\pi}\left(u,\eta'\right)
\, .
\nonumber
\ee
These equations can be easily obtained from \re{loopI0} and  \re{loopI1}
using crossing symmetry. One can investigate all basic properties of ChPT
corrections in the same way as for GPD. We shall not repeat this discussion.

As we can see from the \re{loopDA0} and \re{loopDA1} chiral corrections have
a simple behaviour in parameter $\eta$. Recall that
\be{eta}
\eta={\rm v} \cos\theta_{cm},
\ee
where $\theta_{cm}$ is polar angle of the pion momentum in the CM frame
with respect to the direction of the total momentum {\bf P} and
${\rm v}$ is the velocity of produced pions in the center of mass frame:
\be{gam}
{\rm v}= \sqrt{1-\frac{4m_\pi^2}{\mpp^2}}
\ee
In other words, chiral corrections
contribute only to lowest partial wave
because outgoing pions  can not be produced in state with
higher orbital momentum from the
corresponding diagram  Fig.~\ref{diags}(b).

Let us also note that in the physical region $\mpp>2 m_\pi$ chiral
logarithm generate imaginary part due to two pion intermediate state:
\be{ImI0}
\nonumber
\frac1\pi{\rm Im}\Phi_{2\pi}^{I=0}(u,\eta,\mpp)&=&\theta(\mpp>2m_\pi)
\frac{ [m_\pi^2-2\mpp^2]}{2 (4\pi F_\pi)^2}
\int_{-{\rm v}}^{\rm v} d\eta'
\chil{\Phi}{}^{I=0}_{2\pi}\left(u,\eta'\right)\, ,
\ee
\be{ImI1}
\nonumber
\frac1\pi{\rm Im}\Phi_{2\pi}^{I=1}(u,\eta,\mpp)&=&
\frac{\eta\theta(\mpp>2m_\pi)}{2 (4\pi F_\pi)^2}
\int_{-{\rm v}}^{\rm v} d\eta'
\left[m_\pi^2-(1-\eta'^2)\frac{\mpp^2}{4}\right]
\frac{\partial}{\partial\eta'}
\chil{\Phi}{}^{I=1}_{2\pi}\left(u,\eta'\right)\, ,
\ee
This imaginary part does not depend on the renormalization
scale $\mu_{\chi}$ because to a given accuracy
analytical contributions can not develop
imaginary part of the matrix element.

\section{Chiral corrections to the amplitudes of the hard exclusive processes}
{}From the expressions for the chiral expansion of the pion GPDs and DAs
we can obtain the chiral
expansion of the amplitudes of various  hard exclusive processes.

\subsection{$\gamma^* \gamma\to \pi \pi$ and $\gamma^* N\to 2\pi N'$ near
pions threshold}
We start with the reaction $\gamma^* \gamma\to \pi \pi$ near the threshold
and in the hard regime, {\em i.e.}
the virtuality of the photon $\gamma^*$ is much larger than the typical
hadronic scale $Q\gg \Lambda_{QCD}$. To the leading twist approximation and
 leading order of QCD perturbation theory the amplitude of the reaction is
dominated by the helicity amplitude describing scattering of
transversely polarized  photons \cite{Die99}.
The answer is expressible in terms of the isoscalar
$2\pi$DA  \footnote{For brevity we do not write trivial factors
related to the quark charges}
\be{eq3}
{\cal A}_{++}(\eta,\mpp)=
\int_{-1}^1 du\ \frac{\Phi_{2\pi}^{I=0}(u,\eta,\mpp)}{1-u}\, .
\ee
Substituting here the results for the chiral corrections to the
$2\pi$DA (see eq.~(\ref{loopDA0})) we obtain the chiral expansion of
the corresponding amplitude
\be{loopampl}
{\cal A}_{++}(\eta,\mpp)=
\stackrel{\ o}{\cal A}(\eta)+  \frac{1}{2 (4\pi F_\pi)^2}
[m_\pi^2-2\mpp^2]\
\int_{-1}^1 d\eta'\  \lo\left[m_\pi^2-(1-\eta'^2)\frac{\mpp^2}{4}\right]
\stackrel{\ o}{\cal A}(\eta')
\ee
Here we used an obvious notation $\stackrel{\ o}{\cal A}(\eta)$ for
\be{eq4}
\stackrel{\ o}{\cal A}(\eta)={\cal A}(\eta,\mpp=0)\biggr|_{m_\pi=0}=
\int_{-1}^1 du\ \frac{\chil{\Phi}{}^{I=0}_{2\pi}(u,\eta)}{1-u}\, .
\ee
We note that the amplitude of the hard $\gamma^*\gamma\to\pi\pi$ reactions, in contrast to
its soft counterpart (low photon virtuality), is the same for the $\pi^0\pi^0$ and $\pi^+\pi^-$.
The difference between these final states is the higher twist effect.

{}From eq.~(\ref{loopampl}) we see immediately that the $O(p^2\lo p^2)$
corrections
affect only the S-wave partial amplitude and hence
the whole amplitude is shifted by the $\eta$ independent ``partonic" form
factor ${\cal F}(\mpp)$:

\be{partonicFF}
{\cal F}(\mpp)= \frac{1}{2 (4\pi F_\pi)^2}
[m_\pi^2-2\mpp^2]\
\int_{-1}^1 d\eta'\  \lo\left[m_\pi^2-(1-\eta'^2)\frac{\mpp^2}{4}\right]
\stackrel{\ o}{\cal A}(\eta') \, .
\ee
From \re{ImI0} we obtain that imaginary part of the amplitude is given by
imaginary part  of that formfactor:
\be{Im2pi}
{\rm Im}{\cal A}_{++}=\frac{\pi\theta(\mpp>2 m_\pi)}{2 (4\pi F_\pi)^2}
[m_\pi^2-2\mpp^2]\ \int_{-\gamma}^\gamma d\eta'\ \stackrel{\ o}{\cal A}(\eta')
\ee

At the twist three level there is contribution from the amplitude describing
scattering of longitudinally polarized virtual photon. In the so-called
Wandura-Wilczek approximation the answer can be expressed in terms of the same
isoscalar DA \ci{KM,An}
\be{A01}
{\cal A}_{0+}(\eta,\mpp)=\partial_\eta
\int_{-1}^1 du\ \Phi_{2\pi}^{I=0}(u,\eta,\mpp)\frac{2}{1-u}\ln
\left(1-\frac{1-u}{2}\right)\, .
\ee
The additional derivative in $\eta$ ensures correct symmetrical properties
of the amplitude.
Because of of that derivative the chiral corrections $O(p^2\ln p^2)$
does not contributue to this amplitude and we obtain:
\be{chiexp}
{\cal A}_{0+}(\eta,\mpp)=\partial_\eta
\int_{-1}^1 du\ \stackrel{\ o}\Phi{}_{2\pi}^{I=0}(u,\eta)\frac{2}{1-u}\ln
\left(1-\frac{1-u}{2}\right)\,+ O(p^2) \, .
\ee

In the hard exclusive process of two-pion production off
the nucleon $\gamma^* N\to 2\pi N'$ the dependence of the
amplitude (at the leading order) on the di-pion mass $\mpp$
is governed by the integral (\ref{eq3}) (for the pion pairs in the
$C=+1$ channel) and by the integral:

\be{eqcm3}
{\cal B}(\eta,\mpp)=
\int_{-1}^1 du\ \frac{\Phi_{2\pi}^{I=1}(u,\eta,\mpp)}{1-u}\, ,
\ee
in the $C=-1$ channel \cite{Pol99a,Diehl:1998pg,Lehmann-Dronke:1999aq}.
Substituting here the results for the chiral corrections to the
$2\pi$DA (see eq.~(\ref{loopDA1})) we obtain the chiral expansion of
the corresponding amplitude

\be{cminus}
&&{\cal B}(\eta,\mpp)=
\chil{{\cal B}}(\eta)\
\left(1-\frac{m_\pi^2 \lo m_\pi^2}{(4\pi F_\pi)^2} \right)\\
\nonumber
&& + \frac{ \eta}{2 (4\pi F_\pi)^2}
\int_{-1}^1 d\eta'
\left[m_\pi^2-(1-\eta'^2)\frac{\mpp^2}{4}\right]\ \lo\left[m_\pi^2-(1-\eta'^2)\frac{\mpp^2}{4}\right]
 \frac{\partial}{\partial\eta'}\chil{{\cal B}}\left(\eta'\right)
\, .
\ee

\subsection{ DVCS on the pion target}

The amplitude of the deeply virtual
Compton scattering (DVCS) in the leading order of pQCD is expressible in
terms of the
Compton form factors, which are defined as \cite{Mul94,Ji97b}

\be{eq5}
{\cal H}(\xi,t)=\int_{-1}^1 dx\ \frac{H^{I=0}(x,\xi,t)}{x-\xi+i\varepsilon}\, .
\ee
The imaginary part of the amplitude is defined by the residue at $x=\xi$:
\be{IM}
{\rm Im}{\cal H}(\xi,t)= -\pi H^{I=0}(\xi,\xi,t)\, ,
\ee
and sensitive to the diagonal value of the GPD, therefore it
is not affected by the non-analytic chiral corrections.

Chiral expansion
of the Compton form factors can be written as
\be{eq6}
{\cal H}(\xi,t)&=&\chil{\cal H}{}(\xi)
+  \frac{1}{2(4\pi F_\pi)^2}
[m_\pi^2-2t]\
\int_{-1}^1 d\eta\  \lo\left[m_\pi^2-(1-\eta^2)\frac{t}{4}\right]
\stackrel{\ o}{\cal A}(\eta)\, .
\ee
We see again that the amplitude is shifted by the
$\xi$-independent form factor ${\cal F}(t)$ (see eq.~(\ref{partonicFF}))
in the space-like region.

We want emphasize that the 1-loop chiral corrections affect only the real part
of the Compton amplitude because chiral correction nullifies at the points
$x=\pm \xi$ according to eq.~\re{loopI0}. Hence we see that real and imaginary parts
have a quite different behaviour at the small values of $t$.
This true not only
for isoscalar  GPD which contains the D-term but also for the isovector GPD.
This qualitative difference in $t$-behaviour of the real and the imaginary parts of the
Compton amplitude leads to qualitatively different $t$-behaviour of, say, beam charge and
beam helicity asymmetries. The former is sensitive to the real part and the latter to
the imaginary part of the DVCS amplitude.
Note
the naive factorization ansatz used for modelling of GPD at small $t$
suggests the same $t$-dependence of the real and imaginary parts of the
Compton amplitudes.

\section{Conclusions}
We calculated the leading non-analytic chiral corrections to the generalized parton distributions
and to the two pion distribution amplitudes. In this paper we restricted ourself to the case of the pion.
First we established chiral counting rules for the leading twist operators on the light-cone
 in the effective field theory. On the basis of these counting rules we constructed the corresponding
operators in the effective field theory and further computed the leading non-analytic corrections
to GPDs, $2\pi$DAs and as consequence to the hard exclusive processes like
$\gamma^* \gamma\to\pi\pi$, $\gamma^* N
\to 2\pi N'$, $\gamma^*\pi \to \gamma\pi$, etc.

Obtained results show that the leading non-analytic chiral corrections to the GPDs affect
only in the region $|x|\leq \xi$ \footnote{For isovector GPD there is a $t$-independent correction
``living" on the whole interval of $x$.} and as consequence only the real part of the leading order DVCS
amplitude receives the non-analytical chiral corrections. This implies that the $t$-dependencies of such observables
as charge and helicity beam asymmetries could be qualitatively different even
at small momentum transfer. Hence, the the studying of the $t$-dependence of these
asymmetries can provide us important information about the $t$-dependence of the GPDs.

We checked our results comparing them in the specific limiting cases with known chiral
corrections to form factors and to parton distributions. Additionally we
derived the form of the
parton distributions in the transverse plane at large values of the impact parameter.

Our results for the leading non-analytic chiral corrections to
GPDs and $2\pi$DAs can be used for extrapolation \cite{lattice} of the lattice
and experimental data in $m_\pi$ and $t$. Also the data on the
hard exclusive processes confronted with our chiral perturbation
theory
predictions will be a new, complementary to the low energy soft
reactions, test of the chiral dynamics.

\section{Acknowledgments}
We are thankful to Lech Mankiewicz and Marc Vanderhaeghen for many
valuable discussions. The
work of NK is supported by DFG (project 920585). MVP is supported
by the Sofja Kovalevskaja Programme of the Alexander von Humboldt
Foundation, the Federal Ministry of Education and Research and the
Programme for Investment in the Future of German Government.

\end{document}